\documentclass[twocolumn,preprintnumbers,amssymb,amsmath]{revtex4}
\usepackage{graphicx}% Include figure files 
\usepackage{epstopdf}
\usepackage{dcolumn}% Align table columns on decimal point 
\usepackage{bm}% bold math 
\begin{document} 
\input epsf.tex
\newcommand{\beq}{\begin{eqnarray}}% can be used as {equation} or {eqnarray}
\newcommand{\eeq}{\end{eqnarray}}
\newcommand{\nn}{\nonumber}
\def\ltap{\ \raise.3ex\hbox{$<$\kern-.75em\lower1ex\hbox{$\sim$}}\ }
\def\gtap{\ \raise.3ex\hbox{$>$\kern-.75em\lower1ex\hbox{$\sim$}}\ }
\def\CO{{\cal O}}
\def\CL{{\cal L}}
\def\CM{{\cal M}}
\def\tr{{\rm\ Tr}}
\def\CO{{\cal O}}
\def\CL{{\cal L}}
\def\CM{{\cal M}}
\def\mpl{M_{\rm Pl}}
\newcommand{\bel}[1]{\be\label{#1}}
\def\al{\alpha}
\def\bt{\beta}
\def\eps{\epsilon}
\def\eg{{\it e.g.}}
\def\ie{{\it i.e.}}
\def\mn{{\mu\nu}}
\newcommand{\rep}[1]{{\bf #1}}
\def\be{\begin{equation}}
\def\ee{\end{equation}}
\def\bea{\begin{eqnarray}}
\def\eea{\end{eqnarray}}
\newcommand{\eref}[1]{(\ref{#1})}
\newcommand{\Eref}[1]{Eq.~(\ref{#1})}
\newcommand{\gsim}{ \mathop{}_{\textstyle \sim}^{\textstyle >} }
\newcommand{\lsim}{ \mathop{}_{\textstyle \sim}^{\textstyle <} }
\newcommand{\vev}[1]{ \left\langle {#1} \right\rangle }
\newcommand{\bra}[1]{ \langle {#1} | }
\newcommand{\ket}[1]{ | {#1} \rangle }
\newcommand{\ev}{{\rm eV}}
\newcommand{\kev}{{\rm keV}}
\newcommand{\Mev}{{\rm MeV}}
\newcommand{\gev}{{\rm GeV}}
\newcommand{\tev}{{\rm TeV}}
\newcommand{\mev}{{\rm MeV}}
\newcommand{\mnu}{\ensuremath{m_\nu}}
\newcommand{\mlr}{\ensuremath{m_{lr}}}
\newcommand{\acc}{\ensuremath{{\cal A}}}
\newcommand{\mav}{MaVaNs}

%%%%%%%%%%
%%%%%%%%%%      title page
%%%%%%%%%%

\title{The Status of Inelastic Dark Matter}
\author{David Tucker-Smith}
\email{dtuckers@williams.edu}
\affiliation{Department of Physics, Williams College,  \\
Williamstown, MA 01267, USA
}%
\author{Neal Weiner}
\email{neal.weiner@nyu.edu}
\affiliation{
           Center for Cosmology and Particle Physics,\\
  Dept. of Physics, New York University,\\
New York, NY 10003, USA\\
and\\
Department of Physics, Box 1560, University of Washington,\\
           Seattle, WA 98195-1560, USA\\
}
\date{\today}
\preprint{UW-PT 04/01}
\begin{abstract}
In light of recent positive results from the DAMA experiment, as well as new null results from CDMS Soudan, Edelweiss, ZEPLIN-I and CRESST, we reexamine the framework of inelastic dark matter with a standard halo. In this framework, which was originally introduced to reconcile tensions between CDMS and DAMA, dark matter particles can scatter off of nuclei only by making a transition  to a nearly degenerate state that is roughly $100~\kev$ heavier. We find that  recent data significantly constrains the parameter space of the framework, but that there are still regions consistent with all experimental results.  Due to the enhanced annual modulation and dramatically different energy dependence in this scenario, we emphasize the need for greater information on the dates of data taking, and on the energy distribution of signal and background. We also study the specific case  of ``mixed sneutrino''  dark matter, and isolate regions of parameter space which are cosmologically interesting for that particular model. A significant improvement in limits by heavy target experiments such as ZEPLIN or CRESST should be able to confirm or exclude the inelastic dark matter scenario in the near future. Within the mixed sneutrino model, an elastic scattering signature should be seen at upcoming germanium experiments, including future results from CDMS Soudan.
\end{abstract}
%\keywords{idm}
\maketitle

In light of the WMAP results \cite{Spergel:2003cb}, our understanding of what we do not know has been placed on very strong footing. The overwhelming majority of the energy density of the universe is of unknown origin, with 23\% an unknown dark matter, and 73\% a mysterious dark energy. A determination of the nature of these components would represent  a remarkable advance in our understanding of the universe.
This fact makes the recent DAMA results \cite{Bernabei:2003wy}, which use seven years of data to establish a $6.3 \sigma$ signal consistent with WIMP-nuclei scattering, worthy of particular attention.

One troubling aspect of the DAMA signal has been its apparent disagreement with other experiments, notably CDMS \cite{Akerib:2004fq}, Edelweiss \cite{edelweiss}, ZEPLIN-I \cite{zep1,zep2}, and CRESST \cite{Angloher:2004tr}, when interpreted as evidence of an ordinary WIMP. Efforts to reconcile WIMP search results by modifying the halo profile have met with limited success \cite{Copi:2002hm}, and possibilities such as spin-dependent interactions seem quite constrained from other sources \cite{Ullio:2000bv}. %Indeed, generalized exclusions of DAMA have been presented \cite{Kurylov:2003ra}, although we %shall see these are not model-independent in that they do not apply here.
Indeed, even the generalized analysis of \cite{Kurylov:2003ra} concludes that it is difficult to reconcile the experiments.  However, the class of models considered there does not include the scenario studied here, in which WIMP scattering off of nuclei is dominantly inelastic.

Inelastic dark matter (iDM) \cite{Smith:2001hy} was introduced  to explain the tension between the DAMA four year data \cite{Bernabei:2000qi} and CDMS. One reason DAMA and CDMS are consistent in this framework is that iDM favors heavier target nuclei, such as iodine, over germanium. 
An essential ingredient in testing this framework \cite{Smith:2001hy,Smith:2002af}, is therefore the study of additional heavy target experiments.
Now that we have additional data from heavy target nuclei, at ZEPLIN-I (Xe), and  at CRESST (W), as well as new, stringent limits from CDMS Soudan, it is worth reexamining this scenario to see what parameter space is still allowed.

In the following section, we review the basic features of iDM, what its effects can be on experiments, and possible origins for iDM, namely a heavy Dirac neutrino and a ``mixed'' sneutrino. In section \ref{sec:regions}, we obtain regions of parameter space presently consistent with existing experiments. 
%We will also make model-independent statements about  mixed sneutrino dark matter, regarding the %experimentally allowed regions that are permitted cosmologically, based on relic abundance %calculations.
In section \ref{sec:models}, we investigate whether there are allowed regions consistent with mixed-sneutrino dark matter, based on relic abundance calculations.

\section{Inelastic dark matter}
The iDM scenario features:
\begin{itemize}
\item{A dark matter particle, $\chi_1$, with zero or highly suppressed elastic scattering cross sections off of nuclei.}
\item{A second state, $\chi_2$, heavier than $\chi_1$ by an amount $\delta = m_2-m_1$, which is of the order of a typical halo WIMP kinetic energy.  Generally, we need $\delta \sim 100~\kev$ for weak-scale values of the $\chi_1$ and $\chi_2$ masses.}
\item{An allowed scattering off of nuclei with an inelastic transition of the dark matter particle, i.e., \\ $\chi_1 + n \longrightarrow \chi_2+n$.}
\end{itemize}
Later, we will see that such a peculiar setup can arise naturally, if degenerate states, with elastic scatterings between them, are split by symmetry breaking parameters \cite{Hall:1998ah}. 

The scale of the splitting is an essential feature, because only with $\delta \sim 100~\kev$ can we have interesting effects in terrestrial experiments. For instance, a Bino, with negligible elastic scattering, could in principle scatter into a Higgsino via Higgs exchange, but in this case the splitting is typically far too large (several GeV) for inelastic scattering to be kinematically allowed.
%but we would not consider this to fall generally within the iDM scenario as the splitting, typically %several GeV, makes this kinematically disallowed. 
At the other extreme, a particle with negligible splitting compared to typical kinetic energies would essentially scatter as an ordinary WIMP. In the DAMA analysis of this scenario \cite{Bernabei:2002qr}, this has been referred to, appropriately and descriptively, as ``preferred inelastic scattering.''

\subsection{Consequences of iDM}
Broadly speaking, the iDM scenario  can have three effects on dark matter experiments:
\begin{itemize}
\item{An overall suppression of signal, favoring heavier targets over lighter ones.}
\item{An energy-dependent suppression of signal, suppressing rates of low energy events more than those of high energy events.}
\item{An enhancement of the modulated signal relative to the unmodulated signal.}
\end{itemize}
%At least some of these features presently favor DAMA over existing limits, and, as we will see, allow it %still a great deal of parameter space. 
We will see that these features allow the results of DAMA to be reconciled with the results of other WIMP searches. 
%For now, we will consider these effects in turn.

The central kinematical change is that only those dark matter particles with sufficient incident energy can scatter. This minimum velocity to scatter with a deposited energy $E_R$ is
\be
\beta_{min}=\sqrt{1\over 2 m_N E_R}\left({m_N E_R \over
\mu}+\delta\right),
\label{eq:betamin}
\ee
where $\mu$ is the reduced mass of the iWIMP/target system. In general, there is a broad distribution of velocities in the halo, most commonly considered to be a modified Maxwell-Boltzmann distribution. Therefore, the principle effect at a given experiment is to limit the sensitivity only to a part of the phase space of the halo. 

The first simple observation one can make is that the minimum velocity in eq.~(\ref{eq:betamin}) decreases as one moves to heavier target nuclei. This simple fact alone can reconcile DAMA with light-target experiments, but a full analysis requires us to calculate carefully the event rates at all experiments.

To do this, we follow \cite{Smith:2001hy}.
The differential rate per unit detector
mass is given by 
\begin{equation}
{dR \over dE_R}= N_T {\rho_\chi \over m_\chi} \int_{v_{min}} \!dv \; v
f(v)
{d\sigma \over dE_R}.
\label{eq:rate1}
\end{equation}
Here $N_T$ is the
number of target nuclei per unit mass, $\rho_\chi$ is the local
density of dark matter particles of mass $m_\chi$, ${d\sigma \over
  dE_R}$ is the differential cross section for relic-nucleus
scattering, and $v$ and $f(v)$ are the relic speed and speed
distribution function in the detector rest frame.  We take
$\rho_\chi=.3$ GeV/cm$^3$.

For spin independent interactions, we can write the differential cross section as
\begin{equation}
{d\sigma \over dE_R}={m_N \over 2 v^2}{\sigma_n\over
  \mu_n^2}{\left(f_p Z+f_n(A-Z) \right)^2\over f_n^2}
F^2(E_R),
\label{eq:cs1}
\end{equation}
$\mu_{n}$ is the reduced mass of
the nucleon/WIMP system (not {\em nucleus}/WIMP system), $f_n$ and $f_p$ are the relative coupling
strengths to neutrons and protons, and $\sigma_{n}$ is the
WIMP-neutron cross section at zero momentum transfer, in the elastic $(\delta=0)$ limit.  For consistency, we use the Helm form factor \cite{Engel:1991wq}
\begin{equation}
F^2(E_r)=\left({3 j_1(q r_0)\over q r_0}\right)^2 e^{-s^2q^2},
\end{equation}
with $q=\sqrt{2 m_N E_R}$, $s=1\ \rm fm$, $r_0=\sqrt{r^2-5s^2}$, and
$r=1.2\ A^{1/3}\ \rm fm$.

In the galactic rest frame, we will use the standard Maxwell-Boltzmann distribution of velocities with
 $v_{rms}=\sqrt{3\over 2}v_0$, where  we take $v_0=200$ km/s for
the rotational speed of the local standard of rest. With the recent CDMS Ge result, it is important to be aware of the effect of the finite escape velocity. Details of this cutoff, including its size and the distribution of velocities near it, are very model dependent. As a simple approximation, we will set the differential cross section, as a function of energy, to zero if the minimum velocity exceeds the galactic escape velocity, $\beta_{min}(E_R) > \beta_{esc}$. That is,
 \be
 \frac{d \sigma}{d E_R} \longrightarrow \frac{d \sigma}{d E_R} \Theta(\beta_{esc}-\beta_{min}(E_R)).
\ee
Although this cutoff tends to overestimate the signal when $\beta_{min} \sim \beta_{esc}$, we will not concern ourselves with this here as the details of the galactic cutoff are uncertain. As we will see, the abrupt cutoff produces certain artifacts  in the predicted energy spectra, which
should be interpreted as 
%Rather than interpret these peculiar features as an imprecision in calculating unknown quantities, we take them instead as
signals of cosmological uncertainty.
We choose a relatively high value for the escape velocity, $v_{esc} = 730 {\rm km/s}$ \cite{Kochanek:1995xv}, to obtain the broadest possible region of allowed parameter space.

The Earth moves relative to the galactic rest frame
\begin{equation}
v_e={v_\odot} + v_{orb} \cos\gamma \cos\left(\omega(t-t_0)\right).
\end{equation}
In this expression $v_{orb} =30$ km/s, $\omega =2\pi$/year, $v_\odot=v_0+12$ km/s, 
$t_0 \simeq$ June 2nd, and $\cos \gamma=.51$.  
Taking
dimensionless variables $\eta=v_e/v_0$ and $x_{min}=v_{min}/v_0$, 
performing the velocity integration in eq. \ref{eq:rate1},
and applying the cross section formula in eq. \ref{eq:cs1}, one obtains
\begin{align}
 {dR \over dE_R}= {N_T m_N \rho_\chi \over 4 v_0 m_\chi}F^2(E_R)
{\sigma_n\over
  \mu_n^2}{\left(f_p Z+f_n(A-Z) \right)^2\over f_n^2} \hskip 0.6in \\
 \nonumber \times 
\left({{\rm
  erf}(x_{min}+\eta)-{\rm erf}(x_{min}-\eta)\over \eta}\right)\Theta(\beta_{esc}-\beta_{min}(E_R)).
\label{eq:rate2}
\end{align} 
Notice that the modulation and dependence on $\delta$ are entirely encoded in the second line.

At this point we have obtained all of the results needed to illustrate
%We are now prepared to consider 
the three basic effects that the inelasticity can have at dark matter experiments.

\begin{figure}
\includegraphics[width=3.1 in] {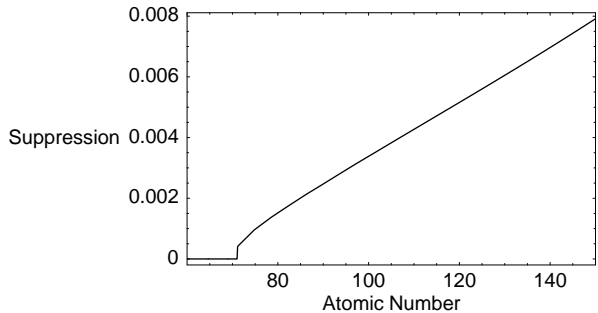}
\caption{\label{fig:supp} The suppression of signal in the energy range  $10 \kev<E_R <150 \kev$ for $m_\chi = 70 \gev$ and $\delta = 100 \kev$ as a function of the atomic number of the target.}
\end{figure}

The simplest of these is the effect on the total rate. Because only a fraction of the velocity space is accessible experimentally, the total rate at an experiment is suppressed considerably. In figure \ref{fig:supp}, we show the dependence of this effect on the atomic number of the target. This illustrates  that heavier targets (such as iodine) can have significantly increased signal over the lighter targets (such as germanium).

\begin{figure}
\includegraphics[width=3.1 in]{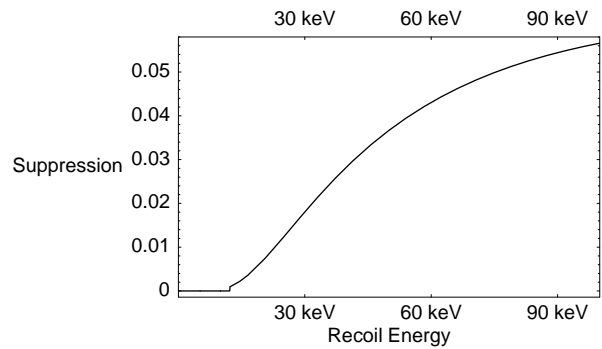}
\caption{\label{fig:endep} The suppression of signal as a function of energy at a germanium experiment, with $m_\chi = 120 \gev$ and $\delta = 80 \kev$.}
\end{figure}

However, it is also important to note where this suppression is coming from. While, indeed, the entire energy range is suppressed, the low energy events are suppressed significantly more. We illustrate this in figure \ref{fig:endep}. This effect tends to favor experiments with higher minimum energies, such as Edelweiss (20 \kev) and DAMA (2 \kev\ with a quenching of .09, or roughly 22 \kev), over experiments with lower thresholds, such as CDMS (10 \kev), ZEPLIN-I (2 \kev\ with a quenching of .2, or roughly 10 \kev), and CRESST (12 \kev).

\begin{figure*}
\centerline{a) \includegraphics[width=3 in] {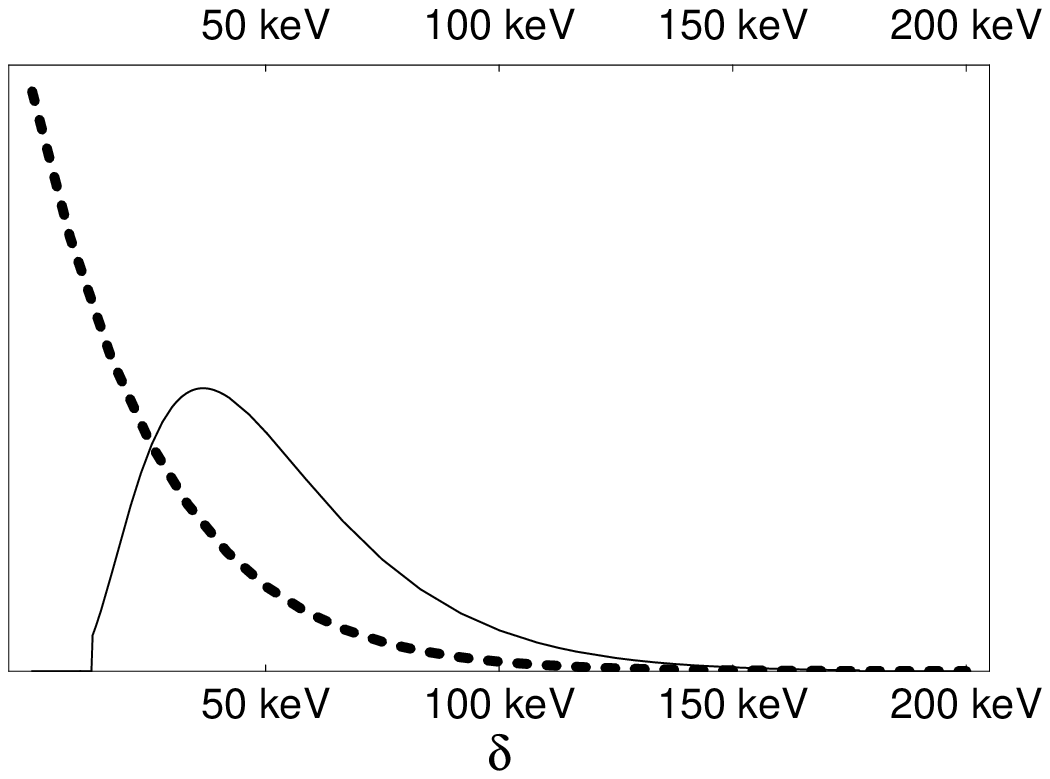} \hskip 0.8 in b) \includegraphics[width=3 in]{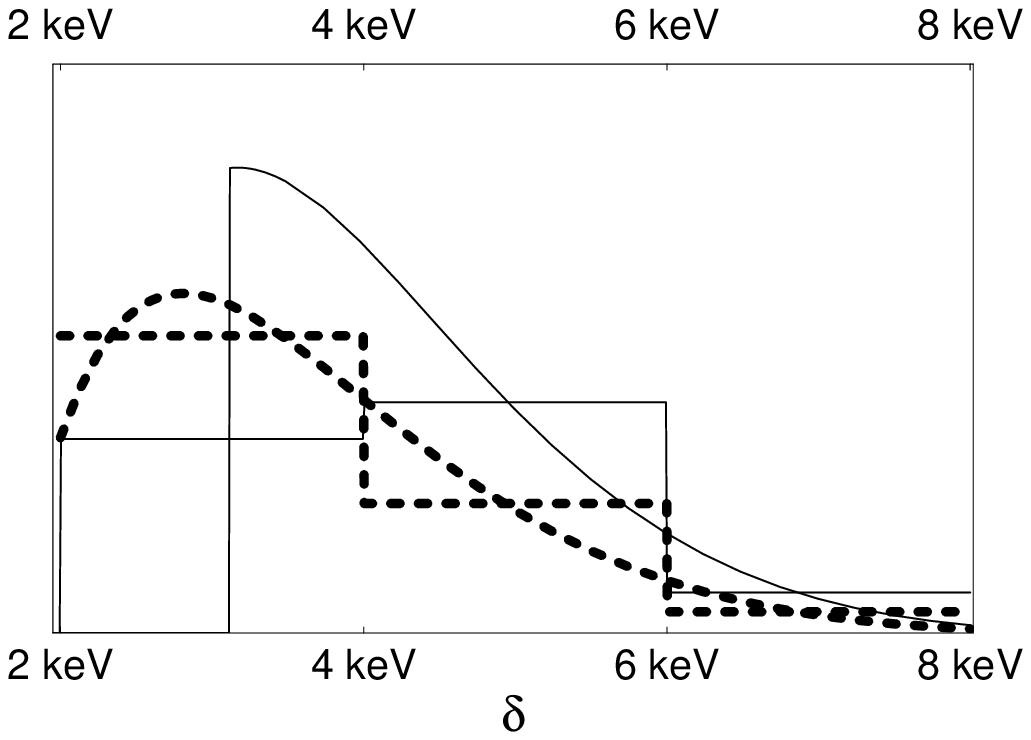}}
\caption{\label{fig:spec}a) The spectrum of signal at a germanium experiment as a function of energy, with area normalized to one. Shown are $m_\chi = 100 \gev$ with $\delta = 80 \kev$ compared with the elastic scattering case. b) Spectrum of the modulated signal at DAMA for $m_\chi = 90 \gev$ and $\delta = 140 \kev$ compared with the elastic case. In both cases, the thin, solid line is the inelastic case, and the dashed, thick line is the elastic case. The sharp cutoff in b) arises due to the finite galactic escape, and would be smoothed with a more realistic cutoff. The histogram shows the integrated signal in the corresponding bins, which is less sensitive to the details of the cutoff.}
\end{figure*}

The energy dependence of the suppression can produce a substantial spectral distortion of the signal.
%Interestingly, a consequence of this is a significant spectral distortion of the signal. 
For instance, while the standard expectation is for the signal to peak at low energies, that need not be the case here, as we show in figures \ref{fig:spec}a) and b).
%there can be significant changes to the spectrum, including a peak at some non-zero energy, both for %modulated and unmodulated components. 
As stated earlier, the abrupt cutoff in the energy spectrum of the inelastic case is an artifact of the sharp cutoff in the signal for velocities above the escape velocity, and should be properly interpreted as a signal of cosmological uncertainties in the spectrum around the cutoff energy. However, there is less uncertainty in the associated histogram, and the zero signal for sufficiently low energies is a robust result.

\begin{figure}[b]
\includegraphics[width=3.2 in]{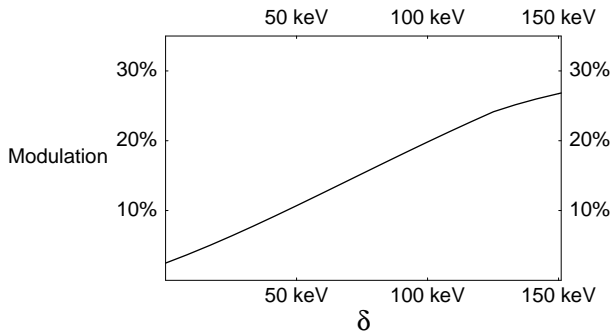}
\caption{\label{fig:mod} Amplitude of modulation as a percentage of the unmodulated signal at DAMA as a function of $\delta$, with $m_\chi=90 \gev$.}
\end{figure}

Finally, there is the enhancement of the modulation signal compared with the unmodulated component. Usually, it is safe to  assume that the modulation will not exceed several percent of the unmodulated signal. However, in the inelastic scenario, we see in figure \ref{fig:mod} that the modulated signal can reach nearly 30\% of the unmodulated signal, improving the comparison of DAMA's modulation result to the unmodulated null results of the other experiments.

Note that the effects of the inelasticity are independent of whether the interactions are spin-independent or spin-dependent.   However, while inelastic spin-dependent interactions are a logical possibility, there is good reason to expect that any  spin-dependent contributions that are present will be elastic (for instance,  in the axial coupling of a fermion, each Majorana component couples to itself, rather than the two components coupling to each other).   For this reason, spin-dependent dark matter probes are not especially sensitive to this scenario. Instead, the essential features of iDM are most directly probed by heavy target experiments, as discussed in section \ref{sec:future}.

\section{Parameter Space}
\label{sec:regions}
Having explored the qualitative changes that arise from iDM,  we can now proceed to a quantitative analysis of the allowed parameter space. It is difficult to  perform a complete analysis as energy dependence is quite important in the iDM scenario, and we lack the full details of the energy spectra obtained experimentally. As such, we will use the following limits for our analysis, which are consistent with published results from DAMA, Edelweiss and CDMS and CRESST, and with preliminary results presented from ZEPLIN-I.

In previous analyses we simply used the DAMA 3$\sigma$ signal in the 2-6 \kev\ region to set our parameters. Now, however, DAMA has given results for both the 2-4 \kev\ region and the 2-6 \kev\ region. We can extrapolate the 4-6 \kev\ region by subtracting off the 2-4 \kev\ signal, and assuming the 2-6 \kev\ error comes from adding the two errors in quadrature. This approach assumes that the systematic effects from the two regions are the same, which is quite reasonable. Ultimately, we take $0.0466\pm0.0094\ \rm cpd/kg$ integrated in the 2-4 \kev\ bins, which is the value given in \cite{Bernabei:2003wy}, and $0.0302\pm0.0081\ \rm cpd/kg$ in the 4-6 \kev\ bins, which is the extracted value.
Using these values, we construct a $\chi^2$ function that depends on the WIMP mass, $\delta$, and $\sigma_n$.  In figures 5 we indicate regions in $\delta$-$\sigma_n$ space with $\chi^2<4,9$ for various WIMP masses.

The advantage of this technique is that we can begin to use the DAMA spectral information to see what regions of parameter space are preferred. 
%This should allow us to have a better sense of what interesting regimes of parameter space exist.  %Likewise, 
DAMA also gives a limit on the maximum oscillation in the higher energy bins of $-0.009 \pm 0.0019\ \rm cpd/kg/\kev$. This places no constraint on the parameter space until very high cross sections ($\sigma_n > 10^{-36}$). 
%We should note that the DAMA experiment has performed their own fit of the iDM scenario %\cite{Bernabei:2002qr, Bernabei:2003wy}, but this analysis gives a parameter space with various %nuclear and halo uncertainties projected onto the parameter space. 
We should note that the DAMA experiment has performed their own fit of the iDM scenario \cite{Bernabei:2002qr, Bernabei:2003wy}, but in this analysis various nuclear and halo uncertainties are projected onto the parameter space. 
As we wish to compare experiments, we work in a single model, without varying the parameters. Earlier analyses had used previous limits from the DAMA pulse shape discrimination, NaI and Xe data. Here, we find this data is subsumed by the ZEPLIN-I limits and do not include it.

For CDMS we require an expectation of fewer than two events in 19.4 kg day of exposure in the energy interval above 10\kev. We do not include Edelweiss at this time since CDMS is the most constraining Ge experiment. ZEPLIN-I has no published results, but has provided preliminary results both at idm 2002 \cite{zep2} and TAUP in 2003 \cite{zep1}. Although detailed limits on how many counts would be allowed do not exist, we can simply normalize our limits to those from ZEPLIN-I at $\delta=0$. Their lowest excluded point lies at approximately $m_\chi \simeq 70\gev$ and $\sigma_n \approx 10^{-42} \ \rm cm^2$. Note that due to the enhanced modulation of this scenario, the sensitivity of the ZEPLIN experiment depends a great deal on the dates of their data taking. The impact can be a factor of two between summer data and winter data. Our limits assume the average, but have approximately 40\% uncertainty for large $\delta$. For CRESST, which is presently background-limited, we use the results  presented in \cite{Angloher:2004tr}, and, as with ZEPLIN, normalize our results to  the limit in the elastic case, which we take to be $\sigma_n < 1.6 \times 10^{-42} \ \rm cm^2$ at $m_{\chi} = 70 \gev$. 

\begin{figure*}
\centerline{a) \includegraphics[width=3 in] {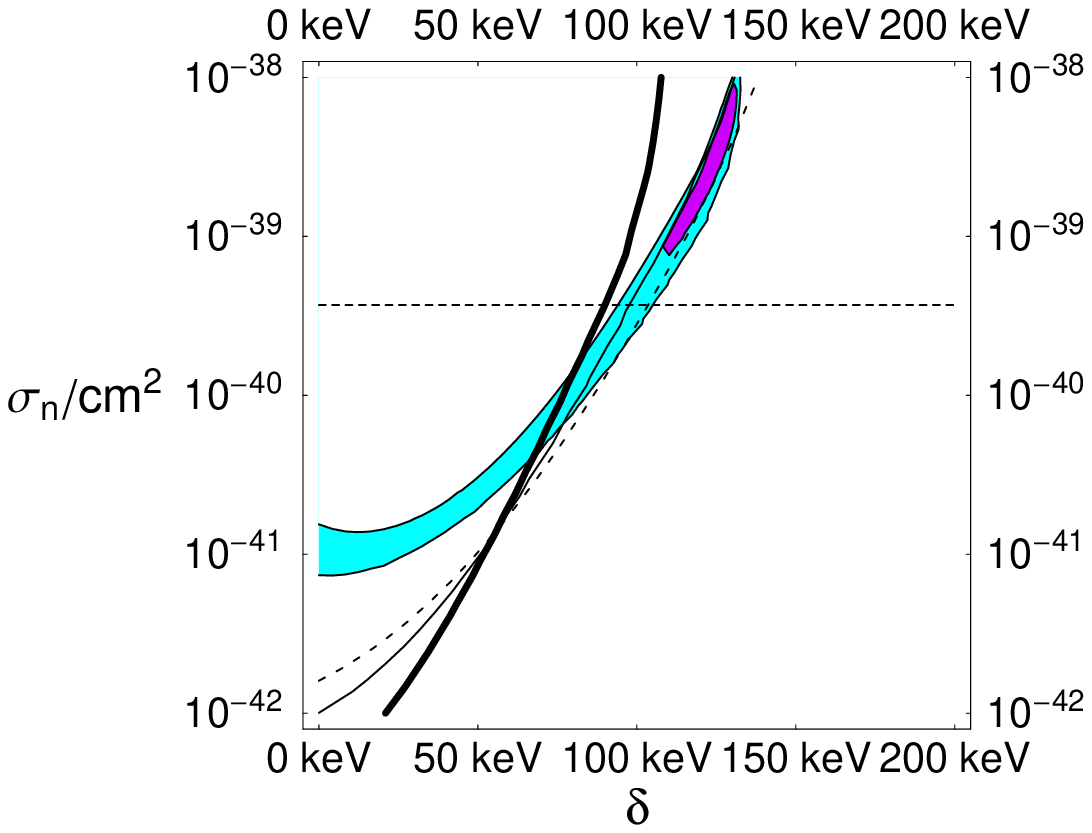}\hskip1 in \epsfxsize=3 in b) \includegraphics[width=3 in] {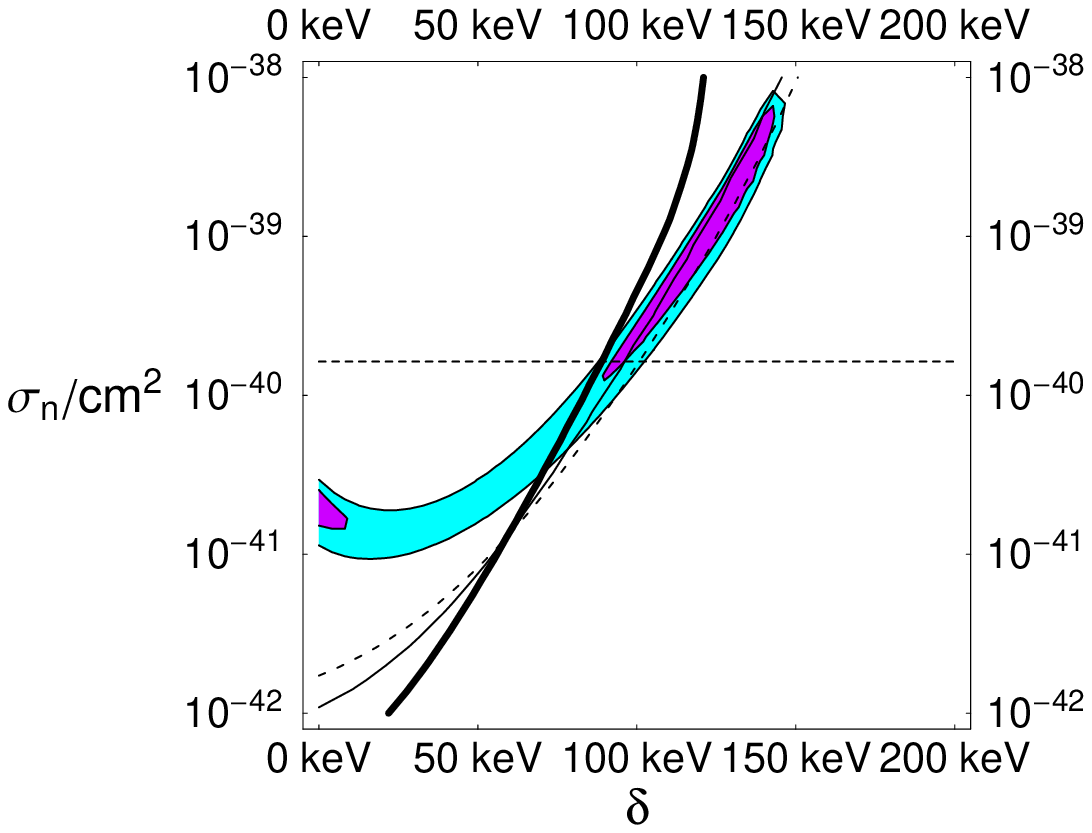}}
\centerline{c) \includegraphics[width=3 in] {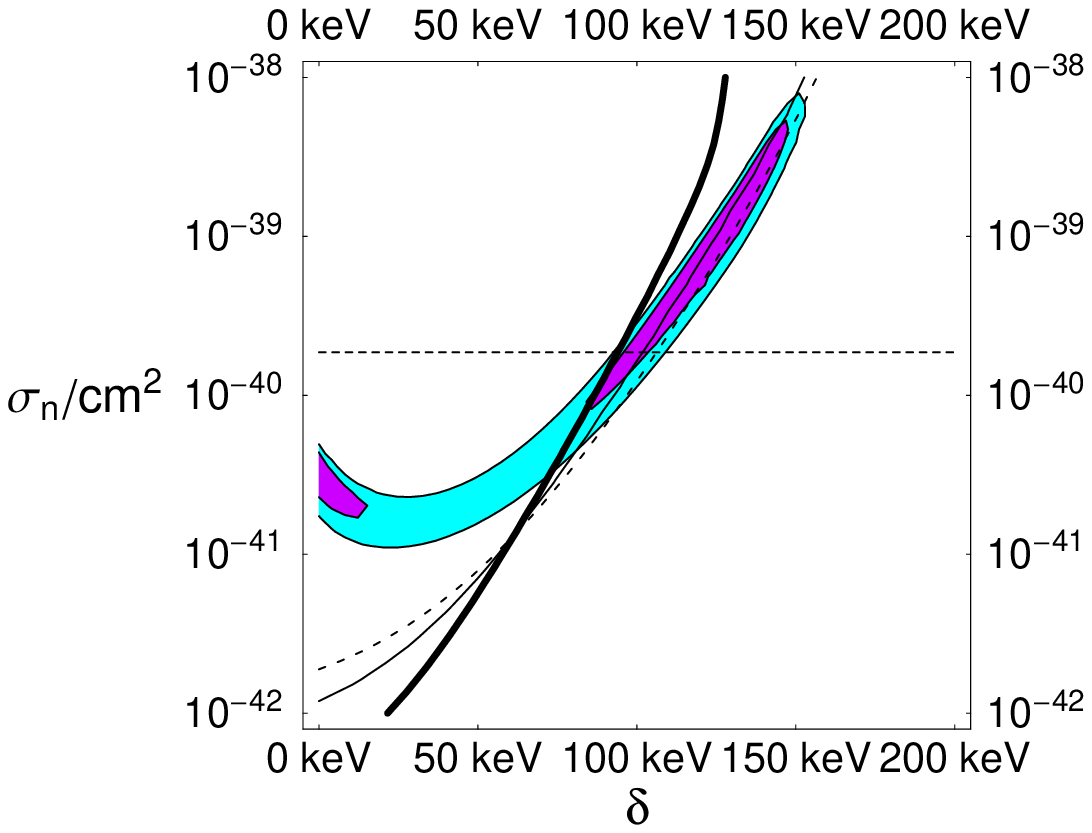}\hskip1 in \epsfxsize=3 in d) \includegraphics[width=3 in] {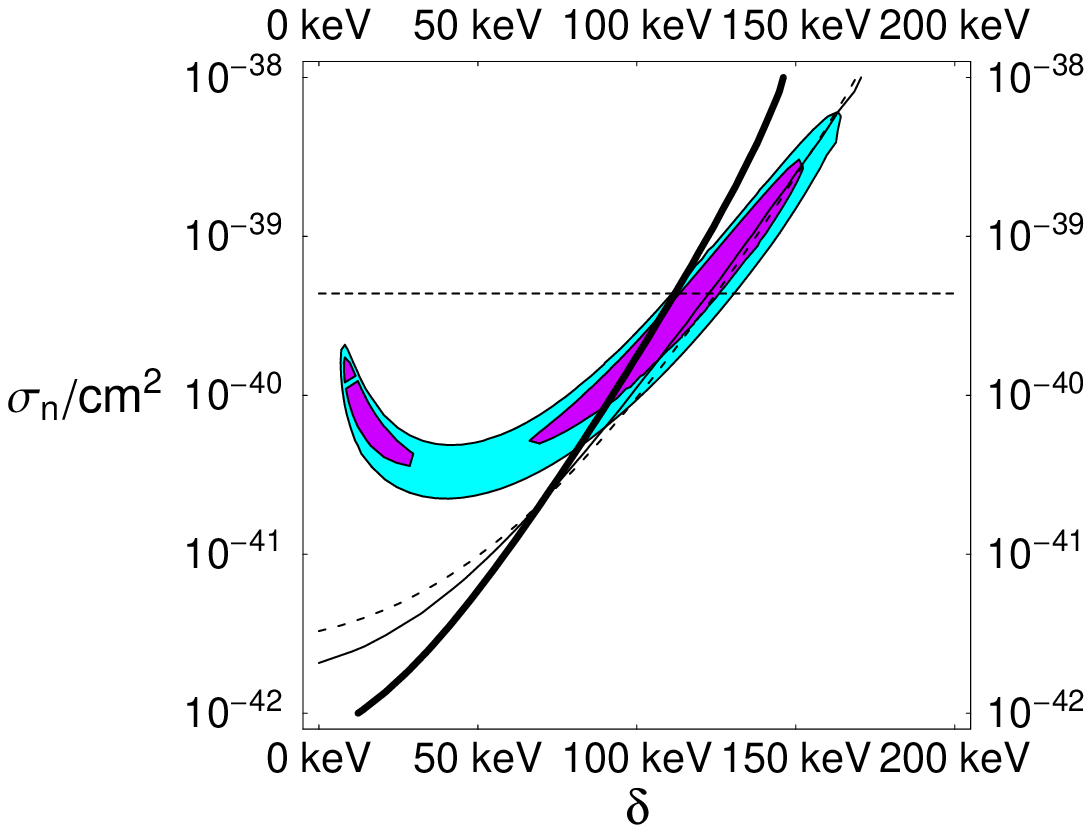}}
\centerline{e) \includegraphics[width=3 in] {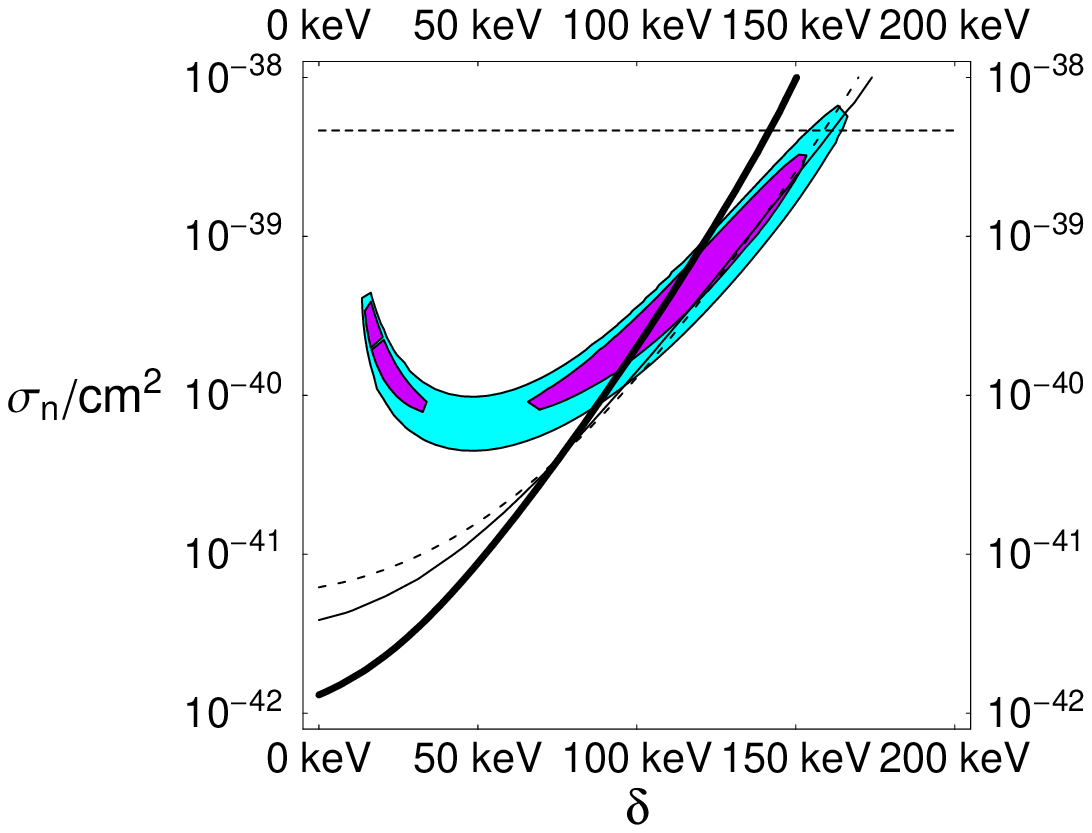}}
\caption{\label{fig:regions} Allowed regions in $\sigma_n$, $\delta$ parameter space for $m_\chi=$ a) 75 \gev\, b) 100 \gev\, c) 120 \gev, d) 250\gev\, e) 500\gev. The light and dark shaded regions have $\chi^2 < 9,4$ as described in the text. The thick solid line is the CDMS limit. The dashed, curved line gives the CRESST results, and the thin solid line gives the preliminary ZEPLIN-I limit. The horizontal dashed line applies to the mixed sneutrino model, and is the upper bound on $\sigma_n$ derived by considering the relic abundance, as described in the text.}
\end{figure*}

We combine these limits in figures \ref{fig:regions} a-e for $m_\chi=75,100,120,250,500$~GeV.  These plots are affected somewhat by value used for $v_0$, which is not a precisely measured quantity.  In general, smaller values of $v_0$ tend to expand the allowed parameter space, and larger values tend to reduce it. These figures should certainly be viewed for their qualitative features primarily, as our use of constraints obtained in the elastic case to determine the constraints for the inelastic case is suspect. In particular, since traditional WIMP signals peak at low energies, an experiment may be able to place strong limits on a WIMP signal, even with higher background at intermediate energies. Since the expected signal for the iDM scenario is in the intermediate energy range, normalizing our limits to the elastic case would then {\em overstate} the limits for the inelastic case. 

Note that the regions preferred by DAMA are disjoint.  For example, for $m_\chi=100$~GeV, there are points with $\chi^2<4$ for very small values of $\delta$,  and  also for large values around $\delta=120$~keV, but not for intermediate values around $\delta=50$~keV.  This is due to the effects of the annual modulation and the inelasticity of the scattering.  For large values of $\delta$, the signal is suppressed at low energies due to the inelasticity.  For very small values of $\delta$, the modulated signal (although not the rate itself), is also suppressed at low energies.  However, at intermediate values of $\delta$, the modulated signal is instead peaked at the minimum energy, 2 keV, so a cross section that leads to consistency with the data in the lower energy bin tends to give too large a modulated signal in the higher energy bin.

Under the assumption that this effect is at most order one, we see from figures \ref{fig:regions} that ZEPLIN and especially CRESST  have placed interesting constraints on the scenario. However, parameter space still exists which is consistent with all experiments simultaneously.  Higher values of $m_\chi$ are somewhat less constrained than lower values.  CDMS, with its new results, is competitive with the heavy target experiments at moderate values of $\delta$, but will not be able to exclude the scenario, even with significantly more data, due to the effect of the finite galactic escape velocity. Later we shall see that CDMS still should see a sub-dominant elastic signal in the case that the iDM particle is a mixed sneutrino. 

\section{Model of inelastic dark matter}
\label{sec:models}
Models of iDM were discussed previously \cite{Smith:2001hy}, but we review the basic approach to model building, and then discuss the ``mixed sneutrino'' case in order to focus our attention to regions of  parameter space where one might expect a relic abundance consistent with observations.

Models of iDM are actually quite simple to construct. We begin by considering the case of a massive Majorana fermion. The simplest way to have a weakly interacting particle scatter off of a nucleus is to have it interact via exchange of a virtual Z-boson. However, massive Majorana fermions do not carry conserved charges, and thus do not have vector couplings to gauge bosons. Yet we know that massive Dirac fermions, such as a fourth generation neutrino, can have such couplings, and a Dirac fermion is nothing more than two Majorana fermions. Hence we can deduce that a vector coupling must allow a transition of one Majorana fermion into another. If the gauge symmetry is broken, these states can be split from one another, and this splitting provides the small inelasticity needed for the framework.

Let us consider the case of fourth generation neutrino more carefully and see how this arises. Take
$\psi=\left(\begin{array}{cc}\eta  & \overline{\xi}\end{array}
\right)$, with vector and axial-vector couplings to quarks:
\begin{equation}
\overline{\psi}\gamma_{\mu}(g_V'+g_A' \gamma_5)\psi \;\overline{q}
\gamma^{\mu}(g_V+g_A \gamma_5) q.
\label{eq:fermioncoupling}
\end{equation}
This term would arise from integrating out massive gauge bosons. The dominant contribution to the scattering of $\psi$ off of
nuclei at a realistic dark matter experiment will come from the vector-vector piece, with an amplitude that scales approximately as the number of nucleons.
The axial-axial piece couples to spin, and has no enhancement from the large number of nucleons.  

Such a fourth-generation neutrino would require a Dirac mass $\sim 100$ GeV for
$\psi$, but after electroweak symmetry breaking, we can also include a small Majorana mass term
${\delta \over 2}(\eta \eta+\overline{\eta}\; \overline{\eta})$, with
$\delta\sim 100$ keV.  The mass eigenstates are Majorana fermions given by
\begin{eqnarray}
\chi_1 \simeq {i \over \sqrt{2}}(\eta-\xi)\hspace{.3in} m_1=m-\delta \\
\chi_2 \simeq {1 \over \sqrt{2}}(\eta+\xi)\hspace{.3in} m_2=m+\delta.
\end{eqnarray}
The vector current, which will dominate if kinematically accessible, couples $\chi_1$ to $\chi_2$, with the elastic scattering coupling suppressed by $\sim \delta /m$:
\begin{equation}
\overline{\psi} \gamma_\mu \psi \simeq i(\overline{\chi}_1
\overline{\sigma}_{\mu} \chi_2-\overline{\chi}_2
\overline{\sigma}_{\mu} \chi_1)+{\delta \over 2 m}(\overline{\chi}_2
\overline{\sigma}_{\mu} \chi_2-\overline{\chi}_1
\overline{\sigma}_{\mu} \chi_1).
\end{equation}
Choosing a Majorana mass splitting of order $100 \kev$,  we have arrived at precisely  the inelastic scenario.

\subsection{Mixed sneutrino iDM}
Although the heavy neutrino provides a nice illustration of how the iDM scenario can be realized, in general its relic abundance is too small. However, in supersymmetric theories, a sneutrino can easily mix through A-terms with a singlet scalar to form ``mixed'' sneutrino dark matter \cite{Arkani-Hamed:2000bq}. Such scenarios can address the origin of neutrino mass \cite{Arkani-Hamed:2000bq,Borzumati:2000mc,Arkani-Hamed:2000kj}, and have a significant impact on collider physics \cite{Chou:2000cy}. Since the lightest mass eigenstate is a linear combination of active and singlet particles, its couplings can be suppressed, and an acceptable relic abundance is easily achieved. We refer the reader to \cite{Arkani-Hamed:2000bq} for details.

Just as we can split a Dirac fermion into two inelastically scattering Majorana fermions, we can split a complex scalar (the lightest mass eigenstate) into two inelastically scattering real scalars. Both possibilities arise from an elastic scattering between degenerate states, which are then split by a small symmetry breaking parameter. This mechanism was first employed in the context of dark matter by \cite{Hall:1998ah}, where the setup of \cite{Grossman:1997is} with an unmixed sneutrino was considered as a possibility for dark matter. 
There, the splitting of scalar and pseudo-scalar states was used to suppress the annihilation rate of sneutrinos in the early universe, so that a cosmologically interesting relic abundance would result.  
%There, the splitting of scalar and pseudo-scalar states both decoupled the heavy state well outside the %region of direct detection experiments and simultaneously changed relic abundances. 
The splitting required for this purpose was quite large, $\delta > 5 \gev$, making such a setup incompatible with the iDM scenario, leaving instead a traditional WIMP scattering via Higgs exchange. However, in the mixed sneutrino setup, the relic abundance is controlled by a mixing angle $\sin \theta$, which specifies what fraction of the light state is active.  Because of this mixing, a value of $\delta$ appropriate for the iDM scenario is viable, and can even arise quite naturally \cite{Arkani-Hamed:2000bq}.

In calculating the relic abundance, contributions to the annihilation rate from Higgs and neutralino exchange depend significantly on details of the model. On the other hand, the interactions with gauge bosons relate directly to the cross section for scattering off of nuclei, and we can obtain upper bounds on the relic abundance by studying the corresponding contributions the annihilation rate.
That is, we obtain robust constraints by considering only annihilation through gauge interactions, while  decoupling the other superpartners and the Higgs.  Below $m_W$, one necessarily has annihilation via s-channel Z to fermions. This is a p-wave interaction, but still requires a $\sin^4 \theta$ suppression to achieve a proper relic abundance. For larger masses, annihilation into $W$ and $Z$ pairs dominates. For sneutrinos near threshold, these channels can significantly suppress relic abundances, while for 
sneutrinos that are 500~GeV or heavier, even $\sin^4 \theta = 1$ is allowed \cite{Falk:1994es}.
This is an interesting point: at about 500\gev\, an ordinary (i.e., unmixed) sneutrino can have the proper relic abundance, and for $\delta\sim 150 \kev$, such a WIMP is apparently still consistent with all data.

For a given sneutrino mass, the requirement that the sneutrino relic abundance be large enough leads to an upper bound on $\sigma_n$.  To calculate this bound we take A-terms of
10 \gev\  and 40 \gev\ , respectively, for sneutrino masses of 120 \gev\ and 250 \gev\ (for lighter sneutrinos the abundance is independent of the A-term; for heavier ones $\sin^2 \theta=1$ is allowed).   
These upper bounds on $\sigma_n$ appear as the horizontal lines in the figures. We have set the relic abundance to $\Omega h^2 = 0.1$ in accordance with various recent results. The overall normalization of the y-axis is uncertain at least to a factor of two, due to the overall uncertainty in the local dark matter density \cite{Kamionkowski:1998xg}, which we incorporate by shifting the sneutrino abundance line up by a factor of two in the cross section space.  Note that for smaller values of $m_\chi$, the relic abundance calculation suggests that there is very little room for mixed-sneutrino iDM, but that the situation improves for larger values of the mass.

\section{Future experiments and elastic scattering signatures}
\label{sec:future}
In the short term, we expect additional results from  xenon, germanium and tungsten experiments. A significant improvement of the sensitivity of the heavy target experiments should certainly confirm or exclude this scenario. The new CDMS results have once again made it competitive with the heavy targets in the moderate $\delta$ range, and continued improvement could exclude this region. However, this depends sensitively on the escape velocity of the galaxy. If it is too low, and $\delta$ is too large there may be no signature at all from inelastic scattering at germanium experiments.

Nonetheless, at least for the sneutrino model we have focused on,  Ge experiments may still detect the WIMP through its {\em elastic} scattering. Although elastic scattering via Z exchange is suppressed, there is still the possibility of scattering via Higgs exchange, with cross section \cite{Arkani-Hamed:2000bq}
\bea
&&  \sigma_n =\\ \nonumber &&\left( \frac{(m_{heavy}^2-m_{\tilde \nu}^2) \sin^2(2\theta) - 2\sqrt{2}m_Z^2 \cos(2 \beta) \sin^2 (\theta)}{200 \gev \times v}\right)^2 \\ \nonumber &&\times \left(\frac{100 \gev}{m_{\tilde \nu}} \right)^2 \left( \frac{115 \gev}{m_h^2}\right)^4 (3 \times 10^{-43} \rm cm^2).
\eea
here, $m_{heavy}^2$ is the mass squared of the heavy linear combination of active and singlet sneutrinos.
%Such a scattering is dependent on an uncertain Higgs-nucleon coupling. 
There is great uncertainty in the strange quark content of the nucleon \cite{Kryjevski:2003mh}, which makes the precise value of the cross section uncertain. Still, upcoming results from CDMS Soudan should be able to test this model, even in the event that $\delta$ is so large that no inelastic scatterings occur.

\section{Conclusions}
We have reexamined the framework of inelastic dark matter in light of  recent data from Edelweiss, DAMA, CDMS Soudan, ZEPLIN-I and CRESST. We find that these experiments have placed interesting constraints on the parameter space of the framework, but that there are still regions which accommodate all experimental results. If $\delta$ is very large, a significant spectrum deformation should be seen in the present DAMA data and in future LIBRA data, but CDMS may see nothing. If $\delta$ is small, CDMS should see spectral distortion, with low energy events suppressed. In all cases, upcoming experiments with heavy targets such as ZEPLIN and CRESST should unambiguously test this scenario

Building models of iDM is quite simple, with fourth generation neutrinos and mixed sneutrinos interesting possibilities. In the sneutrino model, the parameter space is restricted, but still viable. CDMS Soudan should be capable of testing this model through its elastic scattering via Higgs exchange, even in the event that $\delta$ is too large to allow inelastic scattering off of Ge.

\hskip 0.2in

\noindent {\bf Acknowledgments}
The work of NW was partially supported by the DOE under contracts
DOE/ER/40762-213 and DE-FGO3-96-ER40956, and the work of DT-S was 
supported by a Research Corporation Cottrell College Science Award.

\vskip 0.15in
\bibliography{idm05_03_21}
\bibliographystyle{apsrev}

\end{document}